\documentstyle[12pt]{article}
\makeatletter \oddsidemargin  -.1in \evensidemargin -.1in
\newcommand{\singlespacing}{\let\CS=\@currsize\renewcommand{\baselinestretch}{1}\tiny\CS}
\newcommand{\oneandahalfspacing}{\let\CS=\@currsize\renewcommand{\baselinestretch}{1.25}\tiny\CS}
\newcommand{\doublespacing}{\let\CS=\@currsize\renewcommand{\baselinestretch}{1.35}\tiny\CS}

\newtheorem{rule-def}[theorem]{Rule}

\RequirePackage[dvips]{graphicx} \textheight 20.5cm

\begin{document}

\title{\bf Hall effects on MHD free convective flow and mass transfer over a
stretching sheet}
\author{\small G. C. Shit\thanks{Email address: gcs@math.jdvu.ac.in
(G. C. Shit)} \\
\it Department of Mathematics\\
\it Jadavpur University, Kolkata - 700032, India\\
}
\date{}
\maketitle \noindent \doublespacing
\begin{abstract} Of concern in this paper is an investigation of heat and mass
transfer over a stretching sheet under the influence of an applied uniform
magnetic field and the effects of Hall current are taken into account. The non-linear boundary
layer equations together with the boundary conditions are reduced to a system
of non-linear ordinary differential equations by using the similarity
transformation. The system of non-linear ordinary differential equations are
solved by developing a suitable numerical techniques such as finite difference
scheme and Newton's method of linearization. The numerical results concerned with the
velocity, temperature and concentration profiles as well as the skin-friction
coefficient, local Nusselt number $Nu$ and the local sherhood number $Sh$
for various values of the non-dimensional
parameters presented graphically. \\

\end{abstract}

\section {Introduction}
The study of MHD free-convective flow and mass transfer over a stretching
sheet has become a considerable attention to the researchers due to its many
engineering and industrial applications such as, in polymer processing,
electro-chemistry, MHD power generators as well as in flight
magnetohydrodynamics. In the extrusion of a polymer sheet from a die, the
sheet is some times stretched. By drawing such a sheet in a viscous fluid, the
rate of cooling can be controlled and the final product of the desired
characteristics can be achieved. This problem has also an important bearing on
metallurgy where magnetohydrodynamic (MHD) techniques have recently been
used.

Crane (Crane 1970) first introduced the study of steady
two-dimensional boundary layer flow caused by a stretching sheet
whose velocity varies linearly with the distance from a fixed
point in the sheet. Later on several investigators (cf. Gupta and
Gupta 1977; Rajagopal et al. 1984; Siddapa and Abel 1985; Chen and
Char 1988; Laha et al. 1989; Vajravelu and Nayfeh 1992; Sonth et
al. 2002 and Tan et al. 2008) studied various aspects of this
problem, such as the heat, mass and momentum transfer in viscous
flows with or without suction or blowing. The influence of a
uniform transverse magnetic field on the motion of an electrically
conducting fluid past a stretching sheet was investigated by
Pavlov (Pavlov 1974), Chakraborty and Gupta (Chakraborty and Gupta
1979), Kumari et al. (Kumari et al. 1990), Andersson (Andersson
1992, Andersson et al. (Andersson et al. 1992) and Char (Char
1994). The effect of chemical reaction on free-convective flow and
mass transfer of a viscous, incompressible and electrically
conducting fluid over a stretching sheet was investigated by Afify
(Afify 2004) in the presence of transverse magnetic field. In all
these investigations the electrical conductivity of the fluid was
assumed to be uniform. However, in an ionized fluid where the
density is low and/or magnetic field is very strong, the
conductivity normal to the magnetic field is reduced due to the
spiraling of electrons and ions about the magnetic lines of force
before collisions take place and a current induced in a direction
normal to both the electric and magnetic fields. This phenomenon
available in the literature, is known as Hall effect. Thus, the
study of magnetohydrodynamic viscous flows, heat and mass transfer
with Hall currents has important bearing
in the engineering applications. \\

Hall effects on MHD boundary layer flow over a continuous
semi-infinite flat plate moving with a uniform velocity in its own
plane in an incompressible viscous and electrically conducting
fluid in the presence of a uniform transverse magnetic field were
investigated by Watanabe and Pop (Watanabe and Pop 1995).
Aboeldahab and Elbarbary  (Aboeldahab and Elbarbary 2001) studied
the Hall current effects on MHD free-convection flow past a
semi-infinite vertical plate with mass transfer. The effect of
Hall current on the steady magnetohydrodynamics flow of an
electrically conducting, incompressible Burger's fluid between two
parallel electrically insulating infinite planes was studied by
Rana et al. (Rana et al. 2008). \\

The aim of this paper is to study the Hall effects on the steady
MHD free-convective flow and mass transfer over a stretching sheet
in the presence of a uniform transverse magnetic field. The
boundary layer equations are transformed by a similarity
transformation into a system of non-linear ordinary differential
equations and which are solved numerically by using the finite
difference technique. Numerical calculations were performed for
various values of the magnetic parameter, Hall parameter and the
relative effect of chemical diffusion on thermal diffusion
parameters. The results are given for the velocity distribution
and the coefficients of skin-friction along and transverse to the
direction of motion of the stretching surface. Similarity solution
for the temperature field in the above flow is also found and the
rate of heat transfer at the stretching sheet is computed for
various values of magnetic and Hall parameters. Such a study is
also applicable to the elongation to the bubbles and in
bioengineering where the flexible surfaces of the biological
conduits, cells and membranes in living systems are typically
lined or surrounded with fluids which are electrically conducting
(e.g., blood) and
being stretched constantly. \\

\section{Mathematical Formulation}
We consider the steady free-convective flow and mass transfer of
an incompressible, viscous and electrically conducting fluid past
a flat surface which is assuming from a horizontal slit on a
vertical surface and is stretched with a velocity proportional to
distance from a fixed origin O (cf. Fig. 1). We choose a
stationary frame of reference $(x,y,z)$ such that $x$-axis is
along the direction of motion of the stretching surface, $y$-axis
is normal to this surface and z-axis is transverse to the
xy-plane. A uniform magnetic field $B_0$ is imposed along y-axis
and the effect of Hall currents is taken into account. The
temperature and the species concentration are maintained at a
prescribed constant values $T_w$, $C_w$ at the sheet and $T_
\infty$ and $C_ \infty$
are the fixed values far away from the sheet. \\

Taking Hall effects into account the generalized Ohm's law
(cf. Cowling (Cowling 1957)) may be put in the form:\\
\[ J=\frac{\sigma}{1+m^2} \left (E+V\times B-\frac{1}{en_e}J \times B \right ), \]
in which $V$ represents the velocity vector, $E$ is the intensity
vector of the electric field, $B$ is the magnetic induction
vector, $H$ the intensity of the magnetic field vector, $\mu_e$
the magnetic permeability, J the electric current density vector,
$m=\frac{\sigma B_0}{en_e}$ is the Hall parameter, $\sigma $ the
electrical conductivity, e the charge of the electron and $ n_e $
is the number density of the electron. The effect of Hall current
gives rise to a force in the $z$-direction which in turn produces
a cross flow velocity in this direction and thus the flow becomes
three-dimensional. To, simplify the analysis, we assume that the
flow quantities do not vary along z-direction and this will be
valid
if the surface is of very large width along the $z$-direction.\\

Under these assumptions the boundary layer free-convection flow with mass
transfer and generalized Ohm's law is governed by the following system of
equations:

\begin{equation}
\frac{\partial u}{\partial x} + \frac{\partial v}{\partial y} =0
\end{equation}

\begin{eqnarray}
u\frac{\partial u}{\partial x}+v\frac{\partial u}{\partial y}=\nu
\frac{\partial ^2 u}{\partial y^2}+g\beta \left (T-T_\infty\right )+
g\beta^* \left (C-C_\infty\right )-\frac{\sigma B_0^2}{\rho (1+m^2)}\left (
u+mw \right ),
\end{eqnarray}

\begin{eqnarray}
u\frac{\partial w}{\partial x}+v\frac{\partial w}{\partial y}=\nu
\frac{\partial ^2 w}{\partial y^2}+\frac{\sigma B_0^2}{\rho (1+m^2)}\left (
mu-w \right ),
\end{eqnarray}

\begin{eqnarray}
u\frac{\partial T}{\partial x}+v\frac{\partial T}{\partial y}=\alpha
\frac{\partial ^2 T}{\partial y^2},
\end{eqnarray}

\begin{eqnarray}
u\frac{\partial C}{\partial x}+v\frac{\partial C}{\partial y}=D
\frac{\partial ^2 C}{\partial y^2}-k_0\left (C-C_\infty \right)^n,
\end{eqnarray}

where $(u, v, w)$ are the velocity components along the $(x, y, z) $
directions respectively, $\nu $ is the kinematic viscosity, $g $ the
acceleration due to gravity, $\beta $ the coefficient of thermal expansion,
$\beta^* $ the coefficient of expansion with concentration, $T $ and $C $ are
the temperature and concentration respectively, $\rho $ the density of fluid,
$\alpha $ the thermal diffusivity, $D$ the thermal molecular diffusivity,
$k_0$ the reaction rate constant and $n $
the order of reaction. \\

The boundary conditions for this problem can be written as
\begin{eqnarray}
u=bx,~~ v=w=0,~~ T=T_w,~~ C=C_w~~ at ~~y=0
\end{eqnarray}
\begin{eqnarray}
u=w=0,~~ T=T_\infty, ~~ C=C_\infty ~~as ~~ y \rightarrow \infty
\end{eqnarray}
where $b>0$. The boundary conditions on the velocity in (6) are the no-slip
conditions at the surface $y=0$, while the boundary conditions on velocity at
y$\rightarrow \infty $ follow from the fact that there is no flow far way from the
stretching surface. \\
The temperature and species concentration are maintained at a prescribed
constant values $T_w$ and $C_w$ at the sheet and are assume to vanish far way
from the sheet. \\

Introducing the similarity transformation:
\begin{eqnarray}
u=bxf'(\eta), ~~ v=-\sqrt{b\nu} f(\eta), ~~ w=bxg(\eta),
\eta=\sqrt{\frac{b}{\nu}} y \\ \nonumber
\theta (\eta)=\frac{T-T_\infty}{T_w-T_\infty}, ~~ \phi (\eta)=\frac{C-C_\infty}{C_w-C_\infty}
\end{eqnarray}
where a prime denotes derivative with respect to $\eta$.
Substituting (8) into equations (2)-(7), yields
\begin{eqnarray}
f'''+ff''-f'^2+Gr\theta +Gc\phi - \frac{M}{1+m^2}\left (f'+mg\right )=0
\end{eqnarray}
\begin{eqnarray}
g''+fg'-\left (f'+\frac{M}{1+m^2} \right )g+\frac{Mm}{1+m^2}f'=0
\end{eqnarray}
\begin{eqnarray}
\theta''+Prf\theta'=0
\end{eqnarray}
\begin{eqnarray}
\phi''+Sc\left (\phi'f-\gamma \phi^n \right )=0
\end{eqnarray}
where $M=\frac{\sigma B_0^2}{\rho b}$ is the magnetic parameter, \\
$Gr=\frac{g\beta \left (T_w-T_\infty \right )}{b^2x}$ the local Grashof
number, \\
$Gc=\frac{g\beta^*(C_w-C_\infty)}{b^2x}$ the local modified Grashof number, \\
$m=\frac{\sigma B_0}{en_e}$ is the Hall current parameter, \\
$Pr=\frac{\nu}{\alpha}$ the Prandtl number, \\
$Sc=\frac{\nu}{D}$ the Schmidt number, \\
and $\gamma=\frac{k_0}{b}(C_w-C_\infty)^{n-1}$ the non-dimensional chemical
reaction parameter.\\

The boundary conditions (6) and (7) are now obtained from (8) as
\begin{eqnarray}
f'(0)=1, ~~f(0)=g(0)=0, ~~\theta(0)=\phi(0)=1
\end{eqnarray}
\begin{eqnarray}
f'(\infty)=g(\infty)=\theta(\infty)=\phi(\infty)=0.
\end{eqnarray}
The major physical quantities of interest are the skin-friction coefficient
$c_f$, local Nusselt number $Nu$ and the local Sherwood number $Sh$ are
defined respectively by,
 \begin{eqnarray}
C_f=\frac{\tau_w}{\mu\sqrt{\frac{a}{\nu}}ax}=f''(0), ~~where~~ \tau_w=\mu\left
(\frac{\partial u}{\partial y} \right )_{y=0}=\mu\sqrt{\frac{a}{\nu}}axf''(0),
\end{eqnarray}
\begin{eqnarray}
Nu=\frac{q_w}{k\sqrt{\frac{a}{\nu}}(T_w-T_\infty)}=-\theta'(0), ~~where~~ q_w=-k\left
(\frac{\partial T}{\partial y} \right )_{y=0}=-k\sqrt{\frac{a}{\nu}}(T_w-T_\infty)\theta'(0),
\end{eqnarray}
\begin{eqnarray}
Sh=\frac{m_w}{D\sqrt{\frac{a}{\nu}}(C_w-C_\infty)}=-\phi'(0), ~~where~~ m_w=-D\left
(\frac{\partial C}{\partial y} \right )_{y=0}=-D\sqrt{\frac{a}{\nu}}(C_w-C_\infty)\phi'(0),
\end{eqnarray}

When $M=m=0$ and $Gr=Gc=0$, present flow problem becomes
hydrodynamic boundary layer flow past a stretching sheet whose
analytical solution put forwarded by Crane (Crane 1970) as
follows:
\[ f(\eta)=1-e^{-\eta}, ~~~~~ i.e. ~~f'(\eta)=e^{-\eta} \]
An attempt has been made to validate our results for the axial
velocity $f'(\eta)$, we compared our results with this analytical
solution. \\

\section{Numerical Methods}
Several authors Andersson et al. (Andersson et al. 1992), Afify
(Afify 2004) and Abo-Eldahab (Abo-Eldahab 2001) used numerical
techniques for the solution of such two-point boundary value
problems is the Runge-Kutta integration scheme along with the
shooting method. Although this method provides satisfactory
results, it may fail when applied to problems in which the
differential equations are very sensitive to the choice of its
missing initial conditions. Moreover, difficulty arises in the
case in which one end of the range of integration is at infinity.
The end point of integration is usually approximated by replacing
a finite representation to this point and it is obtained by
estimating a value at which the solution will reach
its asymptotic state. \\
On the contrary to the above mentioned numerical method, we used
in the present paper has better stability, simple, accurate and
more efficient numerical technique. The essential features of this
technique is that it is based on a finite difference scheme with
central differencing and based on the iterative procedure. \\

We substitute $F=f'$ and equations (9) and (10) are then
transformed as
\begin{eqnarray}
F''+fF'-F^2+Gr \theta +Gc \phi-\frac{M}{1+m^2}\left (F+mg \right
)=0
\end{eqnarray}
\begin{eqnarray}
g''+fg'-\left (F+\frac{M}{1+m^2} \right )g+\frac{Mm}{1+m^2}F=0
\end{eqnarray}
with the boundary conditions
\begin{eqnarray}
F(0)=1, ~~~F(\infty)=0, ~~~ f(0)=0, ~~~g(0)=0, ~~~ g(\infty)=0
\end{eqnarray}
Central difference scheme is used for derivatives with respect to
$\eta$ :\\
\[ \left (w_\eta \right )_i=\frac{w_{i+1}-w_{i-1}}{2\delta \eta}+
O((\delta \eta)^2) \]
\[ \left (w_{\eta \eta}\right )_i=\frac{w_{i+1}-2w_i+w_{i-1}}{\delta \eta ^2}+
O((\delta \eta)^2) \]

where $w$ stands for $F, ~g,~\theta $ and $\phi$; $i$ is the grid
index in the $\eta$-direction with
\[ \eta_i=i\cdot \delta \eta; ~~~ i=0, 1, 2, \cdots , \]
$\delta \eta $ being the increment along $\eta$-direction. \\
We use Newton's linearization method to linearize the discretised
equations as follows. We assume that the values of the dependent
variables at the $k$th iteration are known. Then the values of the
variables at the next iteration are obtained from the following
equation
\begin{eqnarray}
w_i^{k+1}=w_i^k+ \delta w_i^k
\end{eqnarray}
where $\delta w_i^k$ represents the error at the k-th iteration.
Using (21) in (18) nd (19) and dropping quadratic terms in $\delta
w_i^k$, we get a system of linear algebraic equations for $\delta
w_i^k$. The resulting system of block tri-diagonal equations is
then solved by Thomas algorithm. With an aim to test the accuracy
of this numerical method we have compared the values of $ f'(\eta
)$ for $M=m=Gr=Gc=0$ with those of analytical studies of Crane
(Crane 1970) and have found excellent agreement presented through
the Fig. 2.\\

\section{Results and Discussion}
The system of ordinary differential equations (9)-(12) subject to
the boundary conditions (13) and (14) are solved numerically by
employing a finite difference scheme with Newton's linearization
method described in the previous section. For numerical
computations the following values of the physical parameters have
been considered according to the data used in (Afify 2004): \\
\[ Pr=0.7, ~~ Gr=0.5,~~Gc=0.5, ~~Sc=0.5,
~~n=1,~2,~3;~~M=0~~to~~5.0;~~m=0~~to~~3 ; \]
\[\gamma=0.1,~0.5,~1.0 \]
Figs. 3-5 depicted the variation of velocity profiles $f'(\eta)$,
$f\eta)$ and $g(\eta)$ corresponds to the axial velocity,
transverse velocity and the velocity along z-direction called
cross flow velocity for different values of $M$. From these three
figures we see that the velocity decreases with the increase in
the magnetic parameter $M$. This is to be expected from the
physical consideration since as $M$ increases, Lorentz force which
opposes the flow and leads to deceleration of the fluid motion. It
is interesting to note that for a fixed values of $M$ and $m$,
$f(\eta)$ reaches a uniform value asymptotically at a certain
height $\eta$ above the sheet. By contrast, the cross flow
velocity component induced due to Hall effects and shows a
anomalous behaviour with variation of $M$. Fig. 5 shows that for
fixed values of $M$ and $m$, $g(\eta)$ reaches a maximum value at
a certain height $\eta$ beyond which it decreases gradually in
asymptotic nature. It can be noted from Fig. 5 that in the absence
of magnetic parameter $(M=0)$, cross flow velocity $g(\eta)$
vanishes.\\

Figs. 6 and 7 illustrated distribution of non-dimensional
temperature $\theta(\eta)$ and concentration $\phi (\eta)$ for
different values of magnetic parameter $M$ with a fixed values
$m=0.1$, $Pr=0.7$, $Gr=Gc=0.5$. These two figures shows that the
temperature as well as concentration increases with with the
increase of magnetic parameter $M$. It may conclude that in the
presence of magnetic field temperature and species concentration
can be increased. \\

Effects of Hall current parameter $m$ on the velocity profiles are
presented in Figs. 8-10. For a fixed value of $M$ the velocity
components $f'(\eta)$ and $f(\eta)$ at a given height increases
monotonically with the increase in $m$, while the trend is reverse
in the case of cross-flow velocity $g(\eta)$ induced by Hall
effects. It is interesting to observe from Figs. 8 and 9 that
velocity greatly affected for $m<3$, beyond which no significant
change occur. Figs. 11 and 12 noticed that the temperature and
concentration profiles also affected by the Hall current parameter
$m$. It is shown from Figs. 11 and 12 that the temperature as well
as concentration decreases with the increase in Hall
parameter $m$. \\

The effects of chemical reaction on free-convective flow and mass
transfer of an electrically conducting fluid over a stretching
sheet have also been studied in the presence of magnetic field.
Figs. 13 and 14 shows that velocity and temperature slightly
decreases with the increase of the chemical reaction parameter
$\gamma$, whereas, the concentration significantly diminishes with
an increase in $\gamma$. \\

Fig. 16 shows the variation of skin-friction with $M$ for
different values of Hall parameter $m$. The skin-friction
increases with the increase of both magnetic parameter $M$ and
Hall parameter $m$. The rate of heat transfer at the sheet is
increases with an increase in $m$ but reduces with increasing $M$
shown in Fig. 17. Variation of the rate of mass transfer at the
wall is presented in Fig. 18, which shows that the mass transfer
rate decreases with $M$ but increases with the increase of $m$. It
is interesting to note that the skin-friction coefficient, rate of
heat transfer and mass transfer at the
sheet vary linearly with $M$. \\

\section{Conclusions} The effects of Hall current parameter on free-convective
flow and mass transfer of a viscous, incompressible and
electrically conducting fluid over a stretching sheet have been
studied in the presence of magnetic field. The non-linear boundary
value problem is solved numerically by finite difference scheme
with Newton's linearization method.

From the present investigation, it may be concluded that all the
instantaneous flow characteristics are affected by the Hall
current parameter $m$. In the presence of an external magnetic
field of an electrically conducting fluid, Hall current induced
which in turn produces a cross flow velocity in the direction
perpendicular to the directions of both axial and transverse
velocity. The velocity profiles $f'$, $f$, $g$ decreases with the
increase in $M$ whereas the temperature as well as concentration
increases with the increase of magnetic field strength. The flow
velocities $f'$ and $f$ gradually increases with the increase of
Hall current parameter $m$ while, in the case of cross flow
velocity $g$ and the temperature as well as the
concentration a reversal trend is observed. \\

{\bf Acknowledgement:} {\it The author G. C. Shit is
thankful to the UGC, New Delhi for the
financial support during this investigation. }
\\

\newpage
 \begin{minipage}{1.0\textwidth}
   \begin{center}
      \includegraphics[width=4.0in,height=3.0in ]{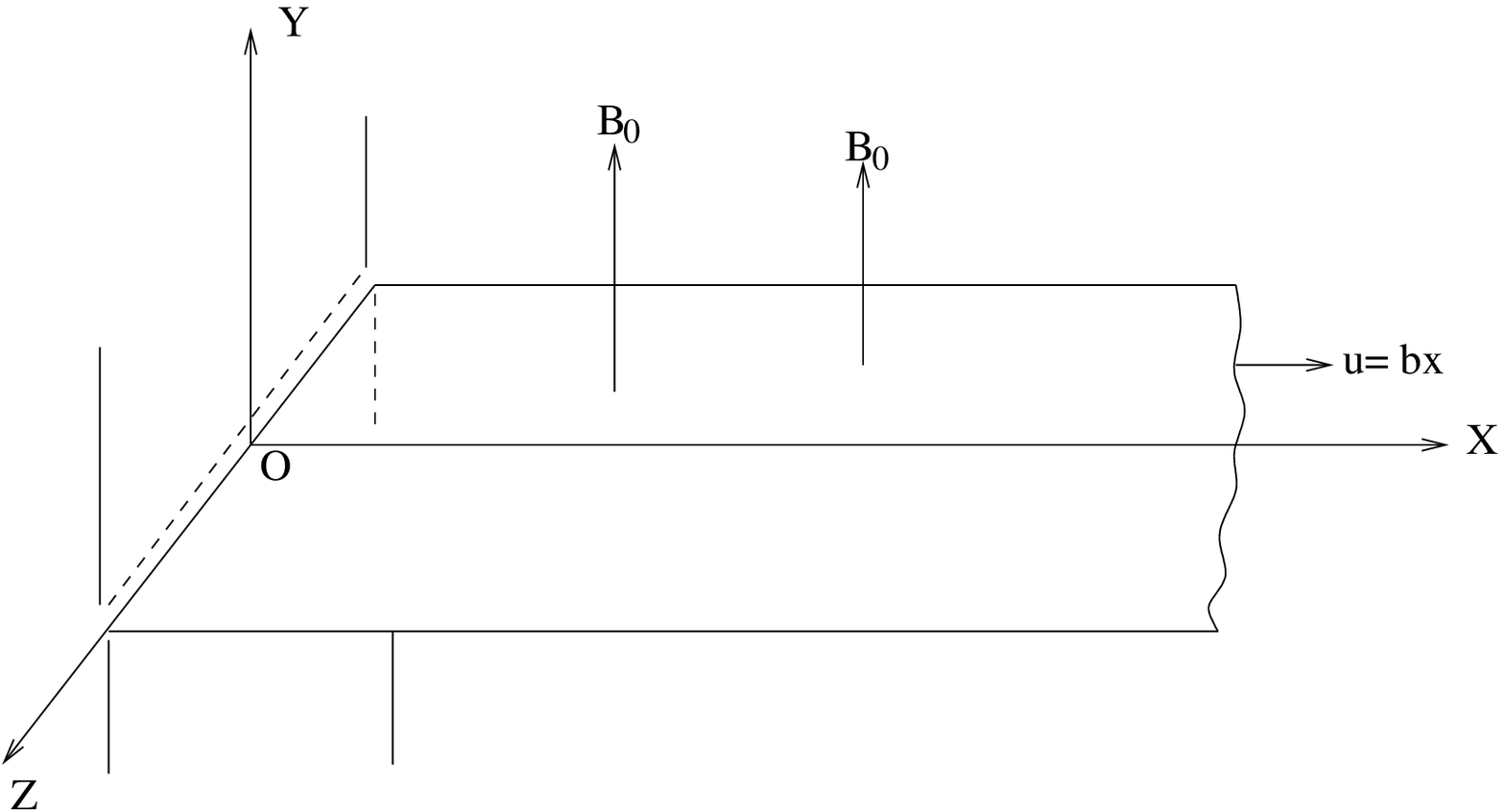}\\
     Fig. 1 Physical sketch of the problem \\

     \includegraphics[width=3.8in,height=3.0in ]{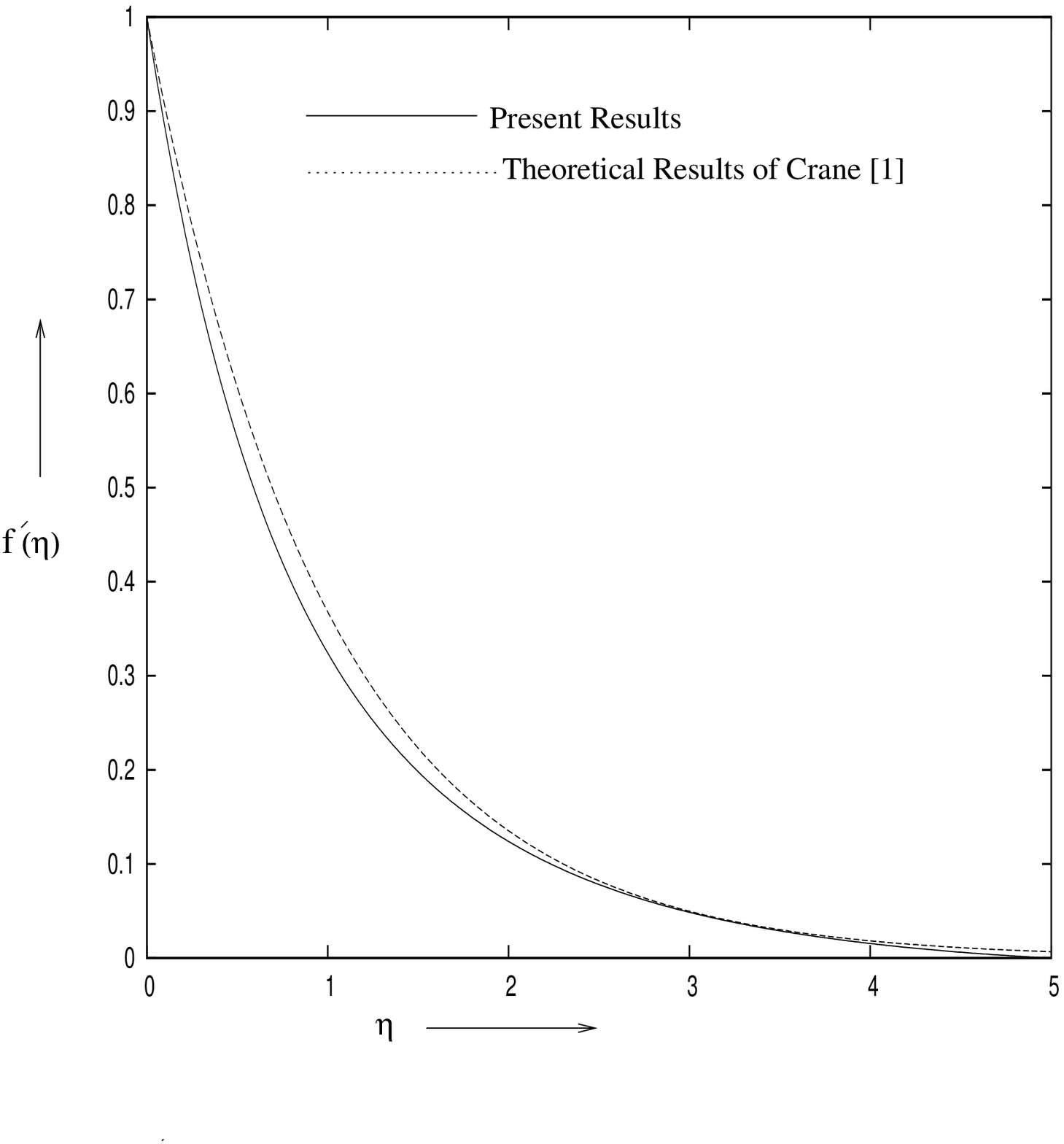}\\
     Fig. 2 Variation of $f'(\eta)$ with $\eta$ for  $M=m=0$
     and $Gr=Gc=0$ \\ (Comparison of the present study with the
     analytical solution of Crane (1970) \\
\end{center}
\end{minipage}\vspace*{.5cm}\\

\begin{minipage}{1.0\textwidth}
   \begin{center}
\includegraphics[width=3.8in,height=3.0in ]{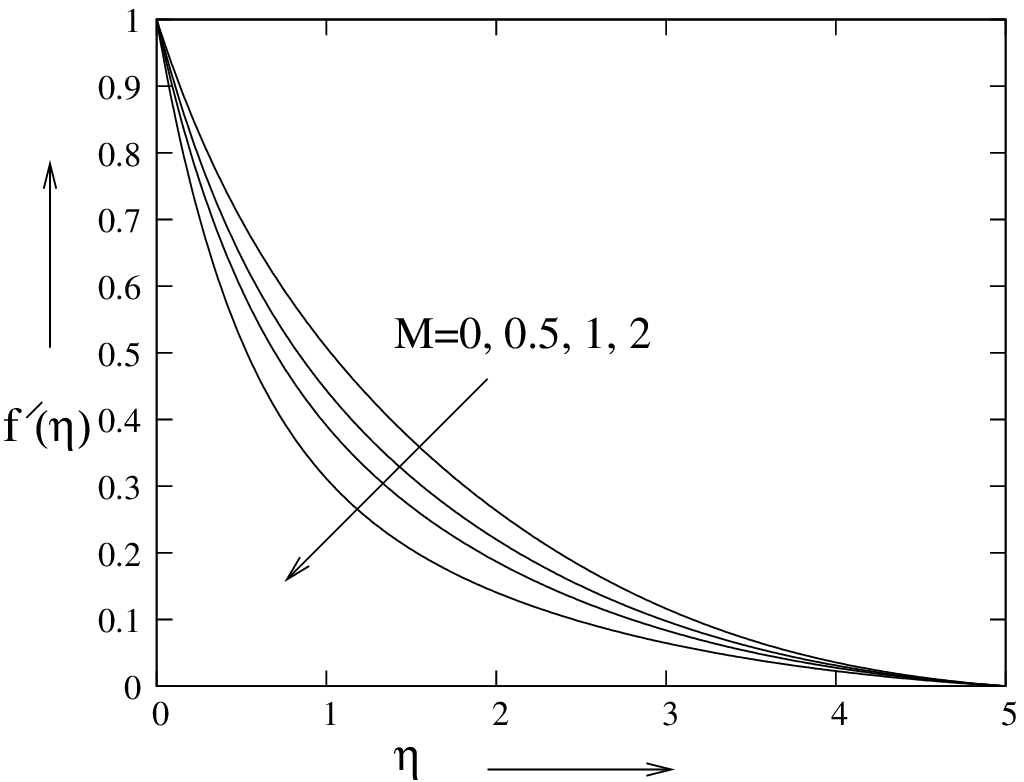}\\
     Fig. 3 Variation of $f'(\eta)$ with $\eta$ for different values of $M$
     and $Pr=0.7, ~Gr=Gc=0.5, ~Sc=0.5, ~n=1, ~m=1, ~\gamma =0.1$ \\
     \includegraphics[width=3.8in,height=3.0in ]{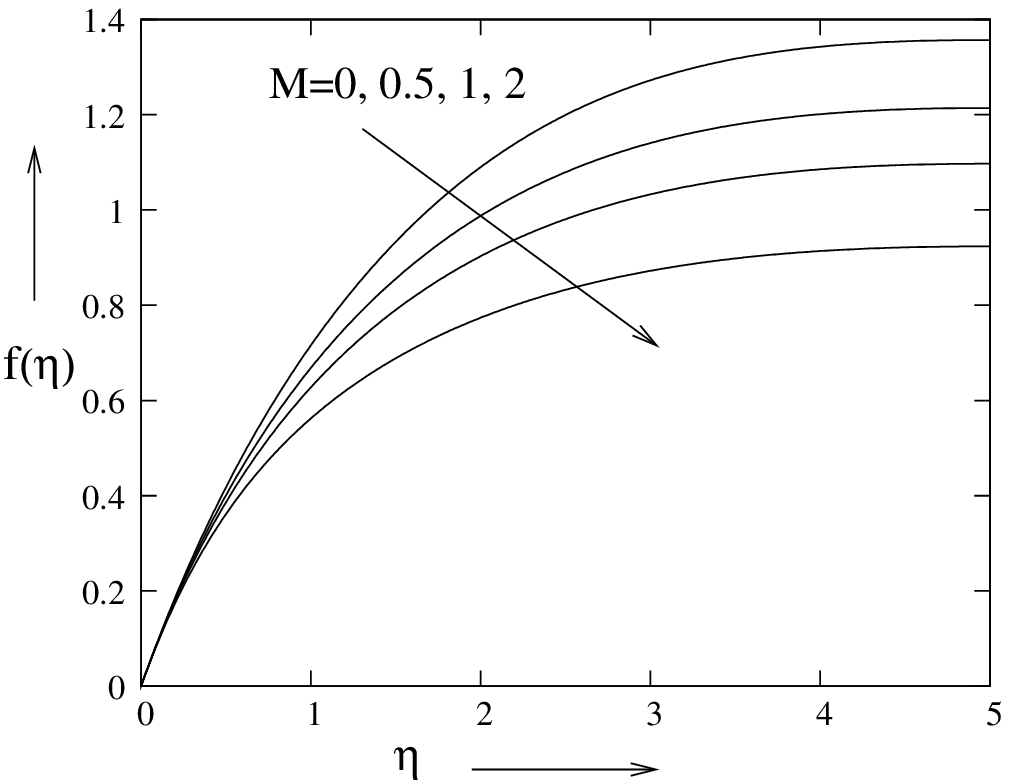}\\

    Fig. 4 Variation of $f(\eta)$ with $\eta$ for different values of $M$
     and $Pr=0.7, ~Gr=Gc=0.5, ~Sc=0.5, ~n=1, ~m=1, ~\gamma =0.1$ \\
\end{center}
\end{minipage}\vspace*{.5cm}\\

 \begin{minipage}{1.0\textwidth}
   \begin{center}
\includegraphics[width=3.8in,height=3.0in ]{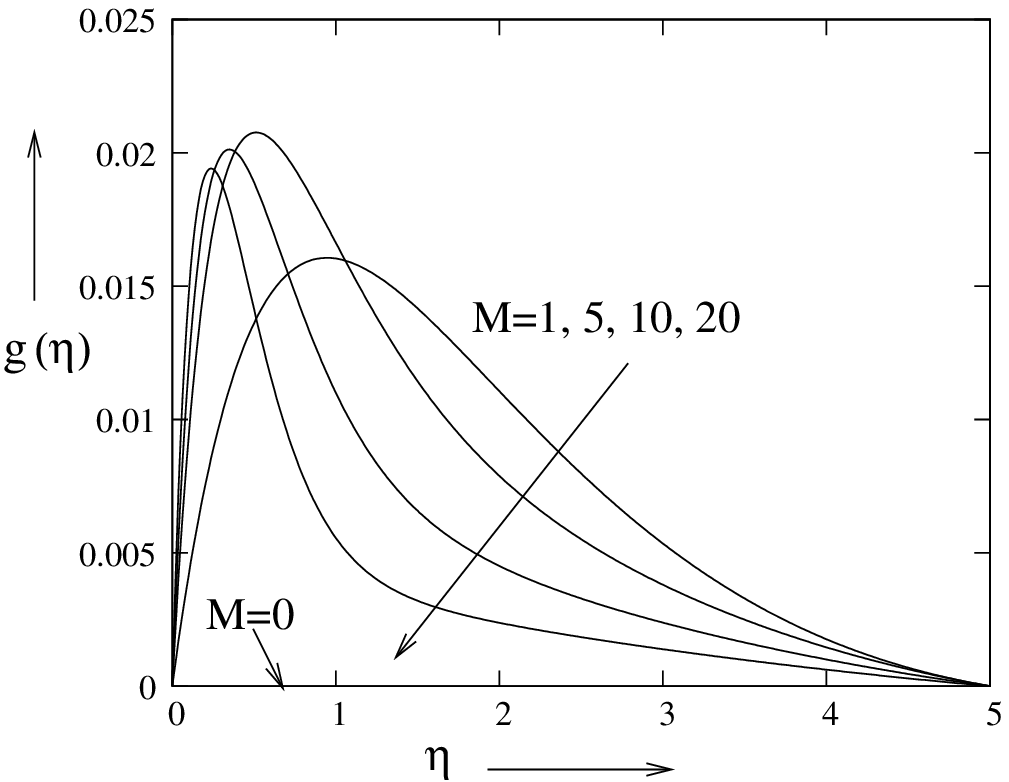}\\
     Fig. 5 Variation of $g(\eta)$ with $\eta$ for different values of $M$
     and $Pr=0.7, ~Gr=Gc=0.5, ~Sc=0.5, ~n=1, ~m=1, ~\gamma =0.1$ \\

     \includegraphics[width=3.8in,height=3.0in ]{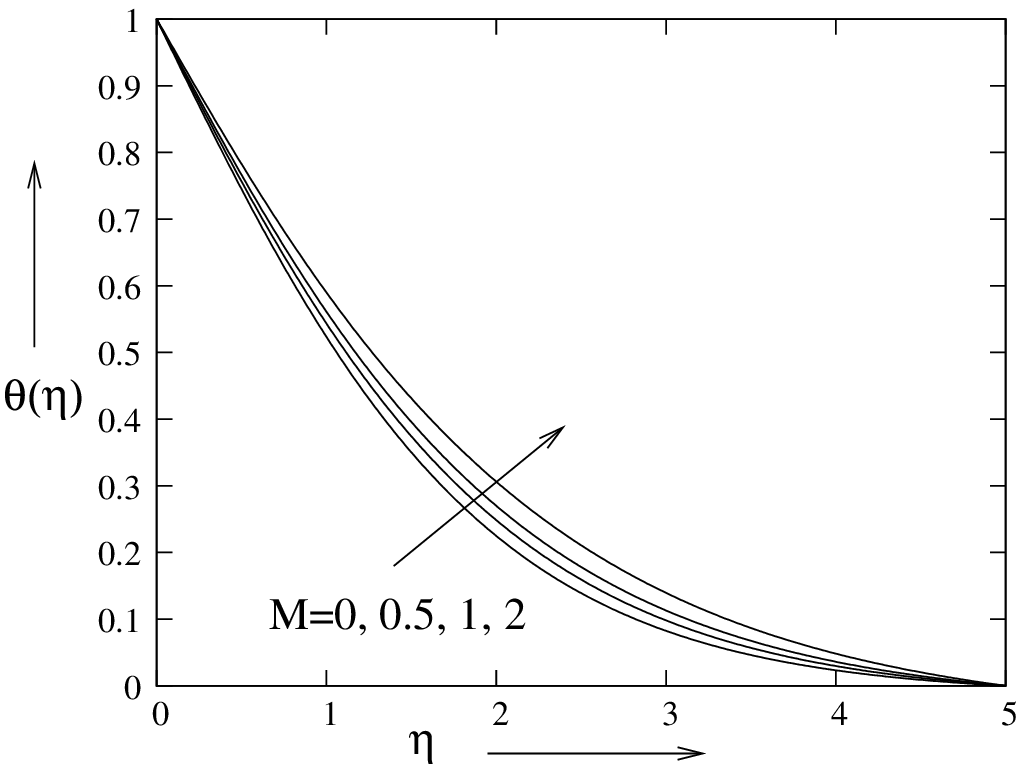}\\
     Fig. 6 Distribution of non-dimensional temperature $\theta(\eta)$ for
     different $M$ with $Pr=0.7, ~Gr=Gc=0.5, ~Sc=0.5, ~n=1, ~m=1, ~\gamma =0.1$ \\
\end{center}
\end{minipage}\vspace*{.5cm}\\

 \begin{minipage}{1.0\textwidth}
   \begin{center}
\includegraphics[width=3.8in,height=3.0in ]{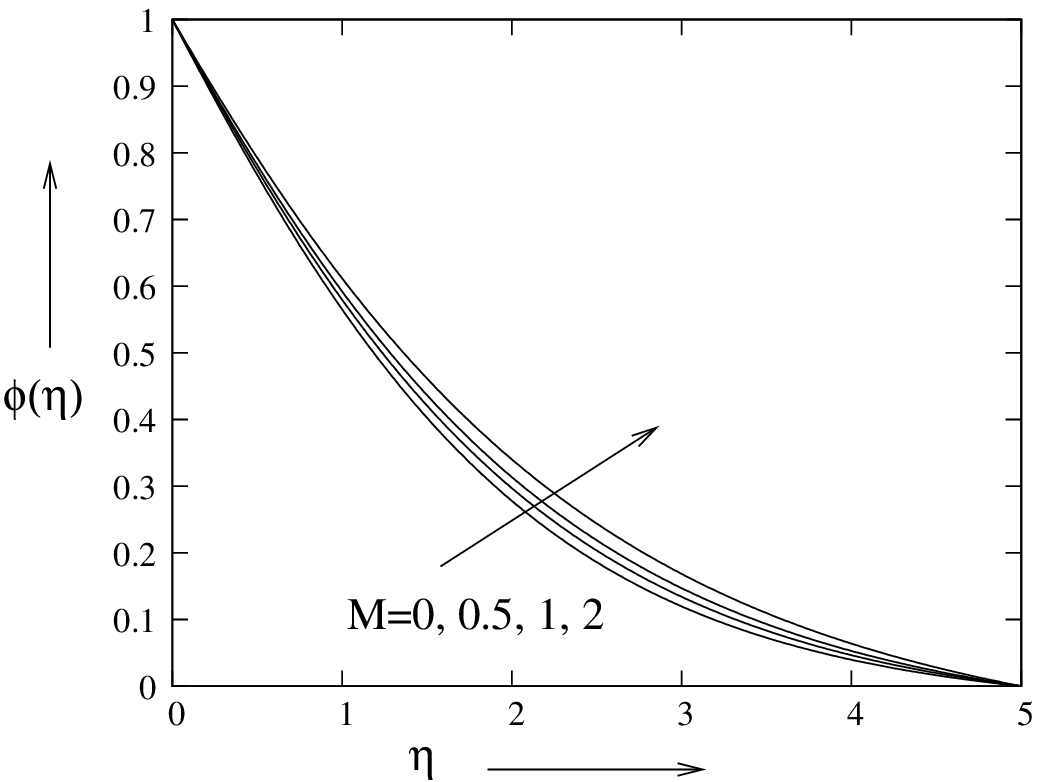}\\
      Fig. 7 The concentration profile $\phi(\eta)$ for
     different $M$ with $Pr=0.7, ~Gr=Gc=0.5, ~Sc=0.5, ~n=1, ~m=1, ~\gamma =0.1$ \\

     \includegraphics[width=3.8in,height=3.0in ]{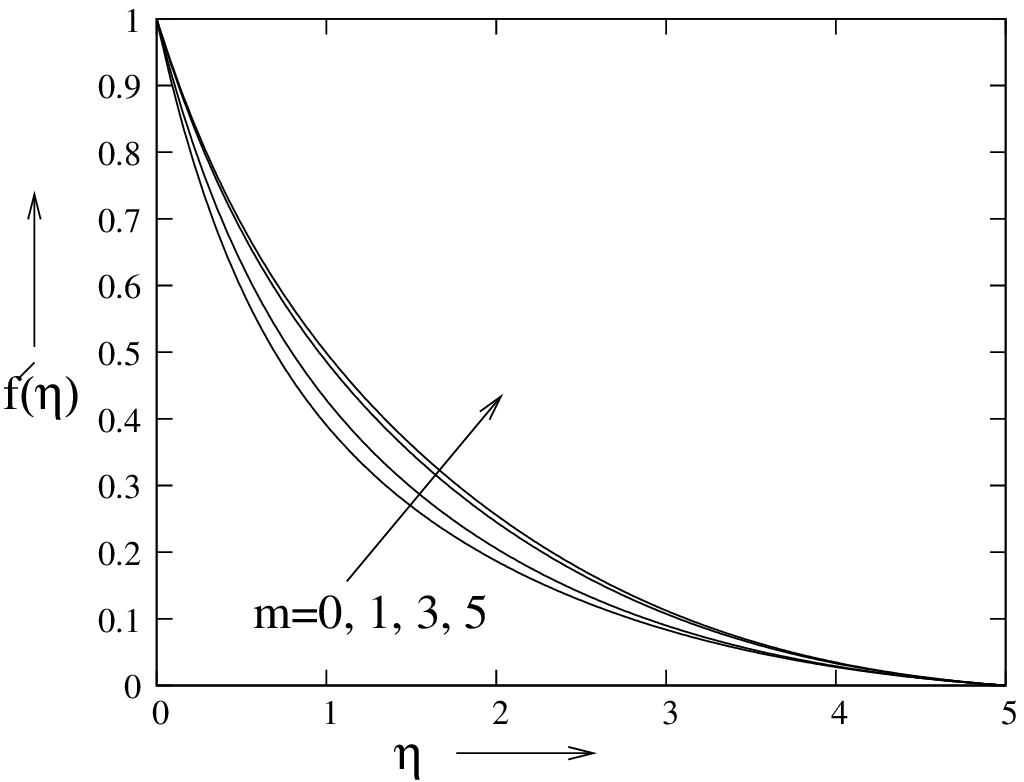}\\
     Fig. 8 Effect of Hall parameter $m$ on axial velocity $f'(\eta)$ for
     $M=1.0, ~Pr=0.7, ~Gr=Gc=0.5, ~Sc=0.5, ~\gamma =0.1, ~n=1$ \\
\end{center}
\end{minipage}\vspace*{.5cm}\\

 \begin{minipage}{1.0\textwidth}
   \begin{center}
\includegraphics[width=3.8in,height=3.0in ]{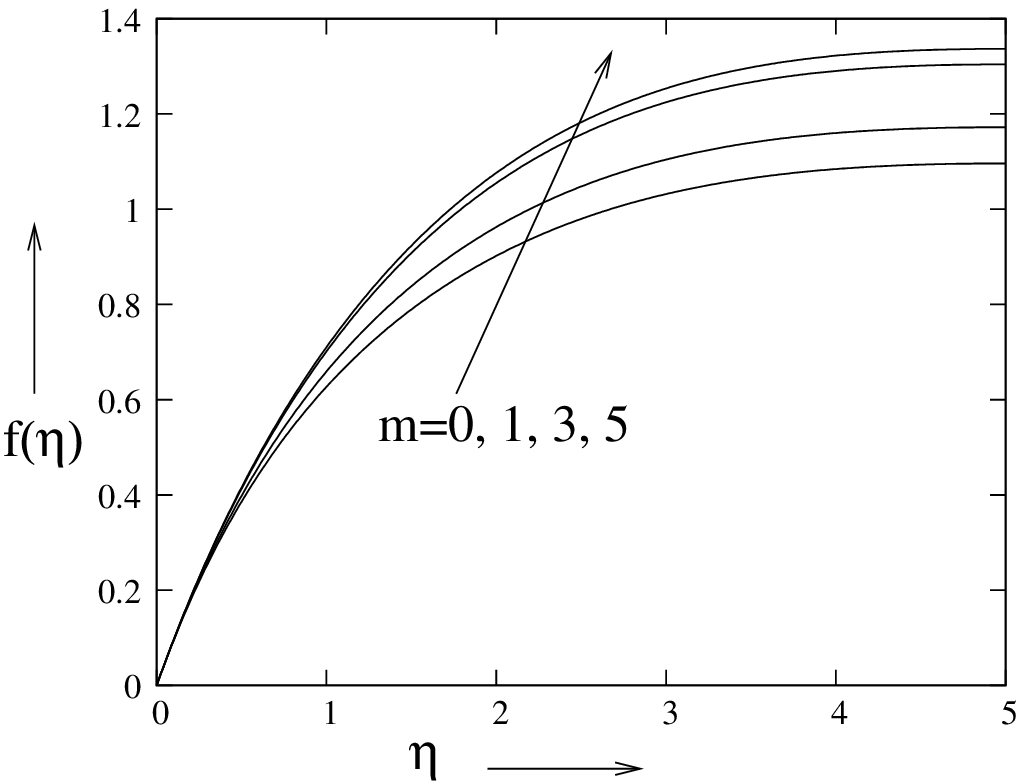}\\
      Fig. 9 Effect of Hall parameter $m$ on axial velocity $f(\eta)$ for
     $M=1.0, ~Pr=0.7, ~Gr=Gc=0.5, ~Sc=0.5, ~\gamma =0.1, ~n=1$ \\

      \includegraphics[width=3.8in,height=3.0in ]{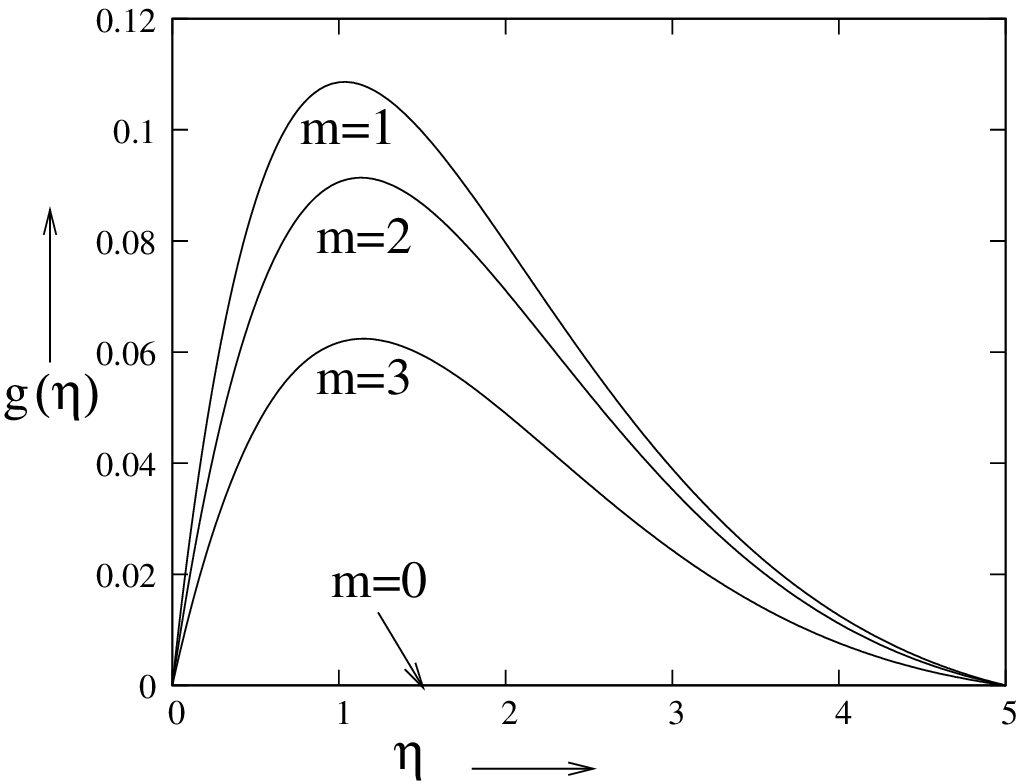}\\
      Fig. 10 Effect of Hall parameter $m$ on the $z$-direction velocity $g(\eta)$ for
     $M=1.0, ~Pr=0.7, ~Gr=Gc=0.5, ~Sc=0.5, ~\gamma =0.1, ~n=1$ \\
\end{center}
\end{minipage}\vspace*{.5cm}\\

 \begin{minipage}{1.0\textwidth}
   \begin{center}
\includegraphics[width=3.8in,height=3.0in ]{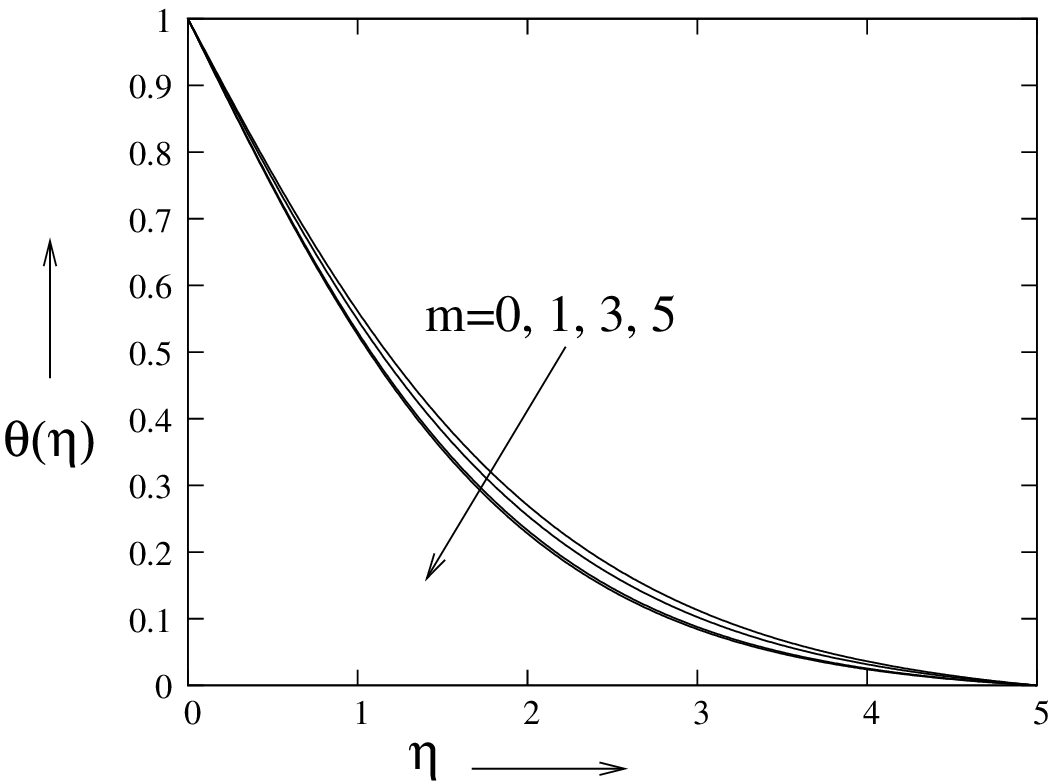}\\
      Fig. 11 Effect of Hall parameter $m$ on non-dimensional temperature $\theta(\eta)$ for
     $M=1.0, ~Pr=0.7, ~Gr=Gc=0.5, ~Sc=0.5, ~\gamma =0.1, ~n=1$ \\

      \includegraphics[width=3.8in,height=3.0in ]{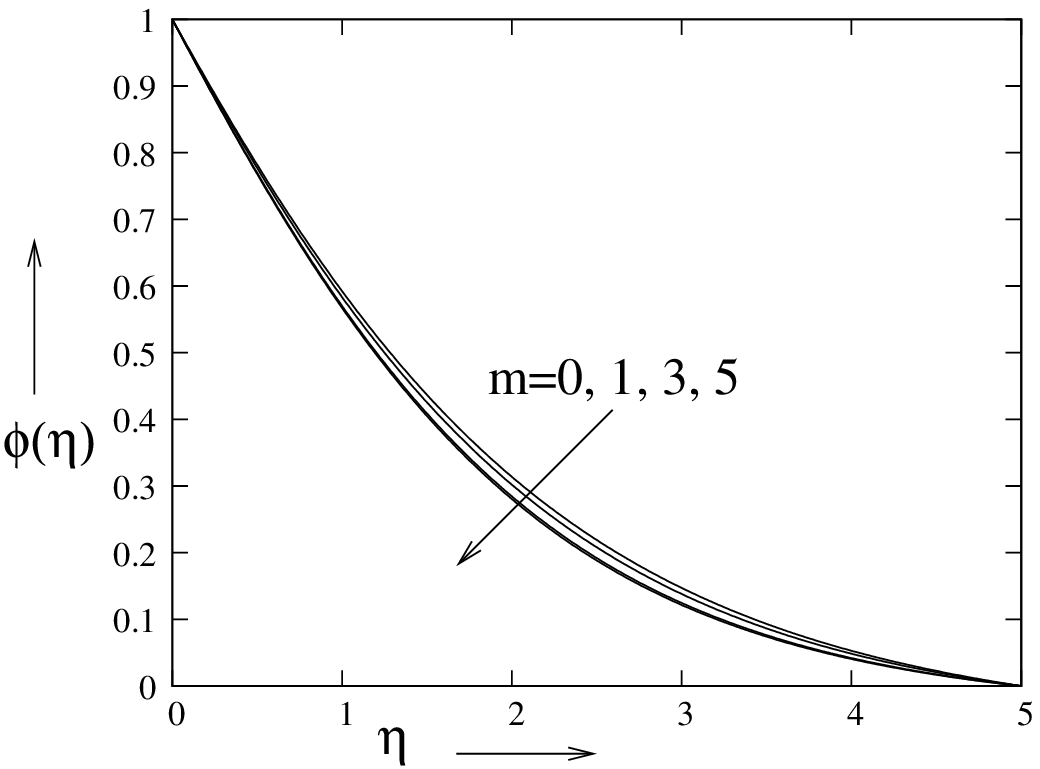}\\
     Fig. 12 Effect of Hall parameter $m$ on the concentration profile $\theta(\eta)$ for
     $M=1.0, ~Pr=0.7, ~Gr=Gc=0.5, ~Sc=0.5, ~\gamma =0.1, ~n=1$ \\
\end{center}
\end{minipage}\vspace*{.5cm}\\

 \begin{minipage}{1.0\textwidth}
   \begin{center}
\includegraphics[width=3.8in,height=3.0in ]{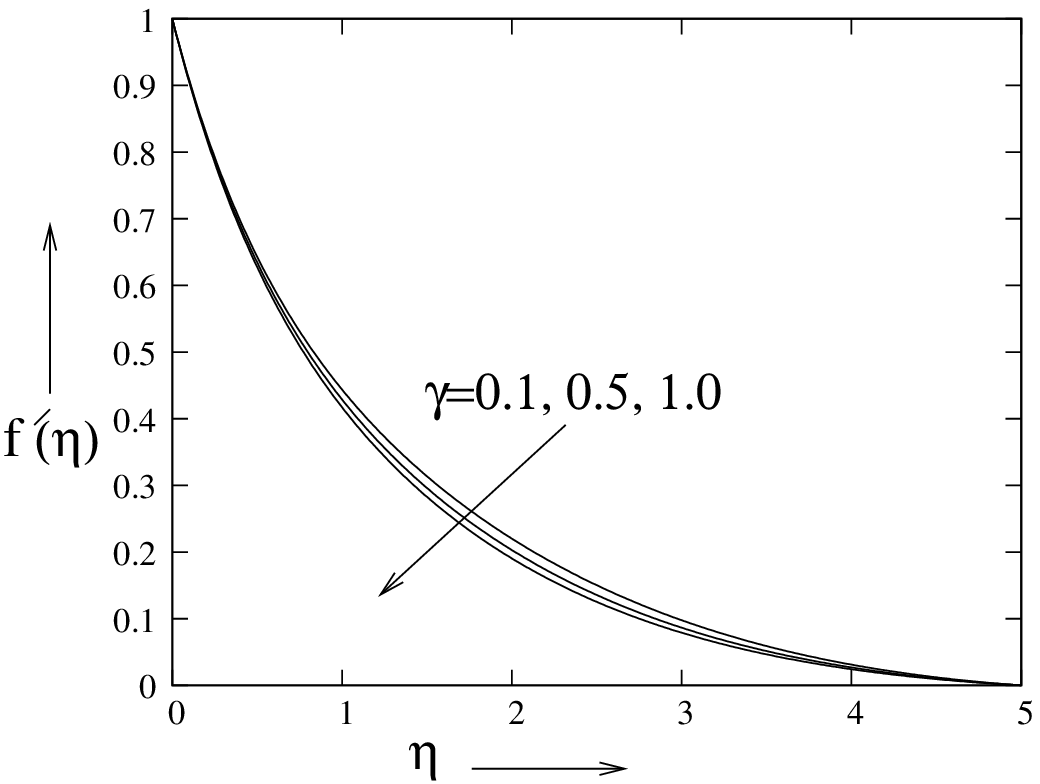}\\
     Fig. 13 Variation of axial velocity profile $f'(\eta)$ with $\eta$ for
     different $\gamma $ and $Pr=0.7, ~Gr=Gc=0.5, ~Sc=0.5, ~M=0.5, ~m=0.1, ~n=1$ \\

     \includegraphics[width=3.8in,height=3.0in ]{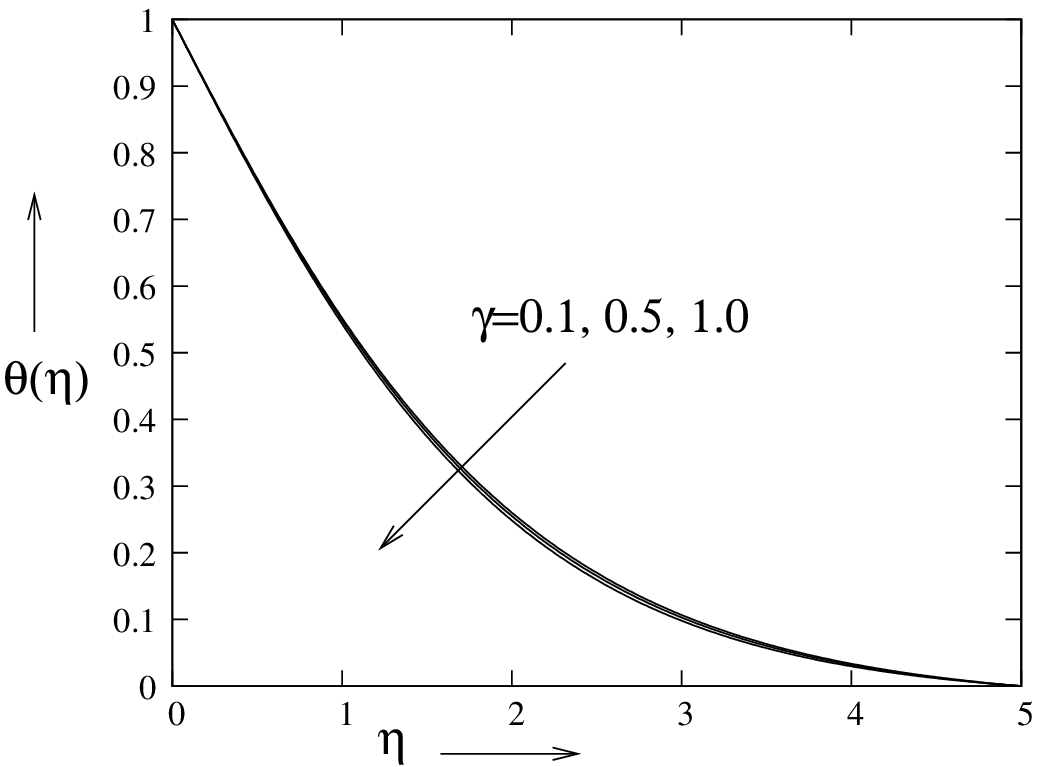}\\
     Fig. 14 Variation of the temperature profile $\theta(\eta)$ with $\eta$ for
     different $\gamma $ and $Pr=0.7, ~Gr=Gc=0.5, ~Sc=0.5, ~M=0.5, ~m=0.1,
     ~n=1$ \\
\end{center}
\end{minipage}\vspace*{.5cm}\\

 \begin{minipage}{1.0\textwidth}
   \begin{center}
\includegraphics[width=3.8in,height=3.0in ]{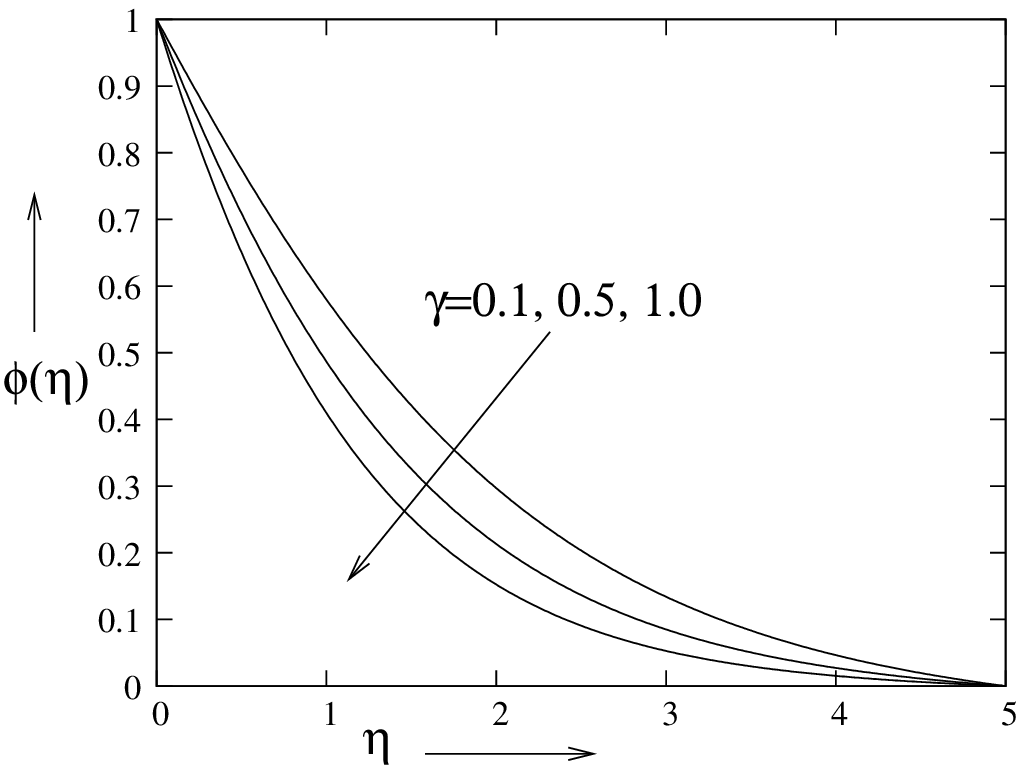}\\
     Fig. 15 Variation of the concentration profile $\phi(\eta)$ for
     different $\gamma $ and $Pr=0.7, ~Gr=Gc=0.5, ~Sc=0.5, ~M=0.5, ~m=0.1, ~n=1$ \\

     \includegraphics[width=3.8in,height=3.0in ]{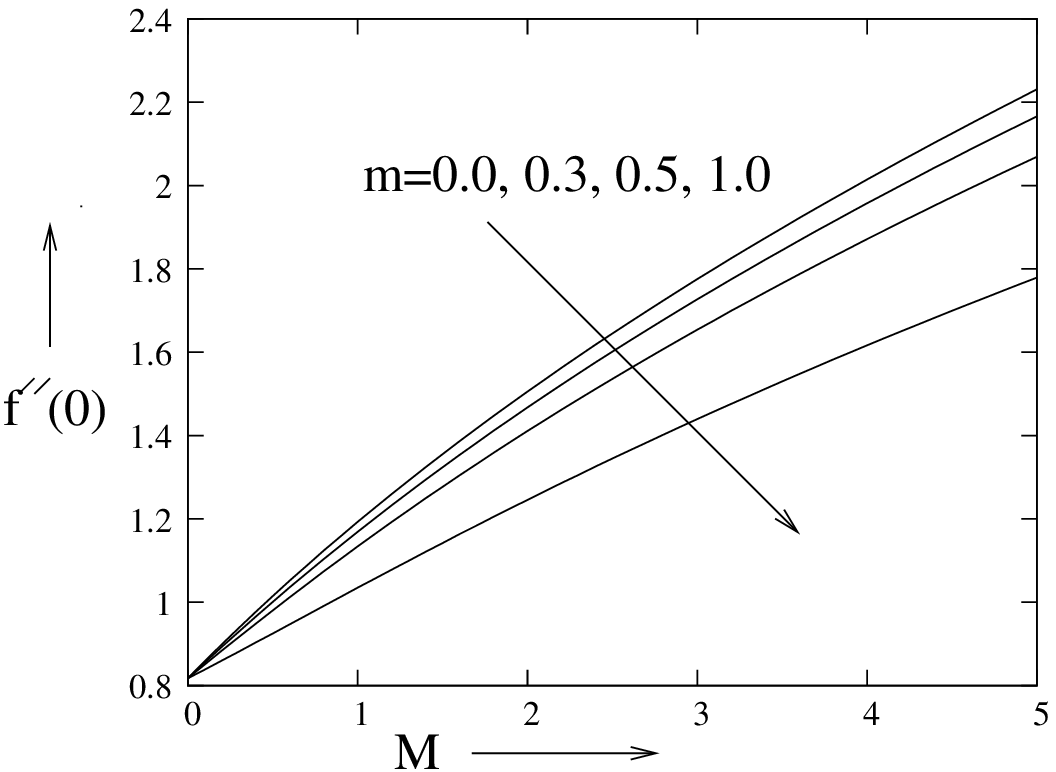}\\
     Fig. 16 Variation of skin-friction for different values of Hall parameter
     $m$ and $Pr=0.7,~ Gr=Gc=0.5, ~Sc=0.5, ~n=1, ~\gamma =0.1 $ \\
\end{center}
\end{minipage}\vspace*{.5cm}\\

 \begin{minipage}{1.0\textwidth}
   \begin{center}
\includegraphics[width=3.8in,height=3.0in ]{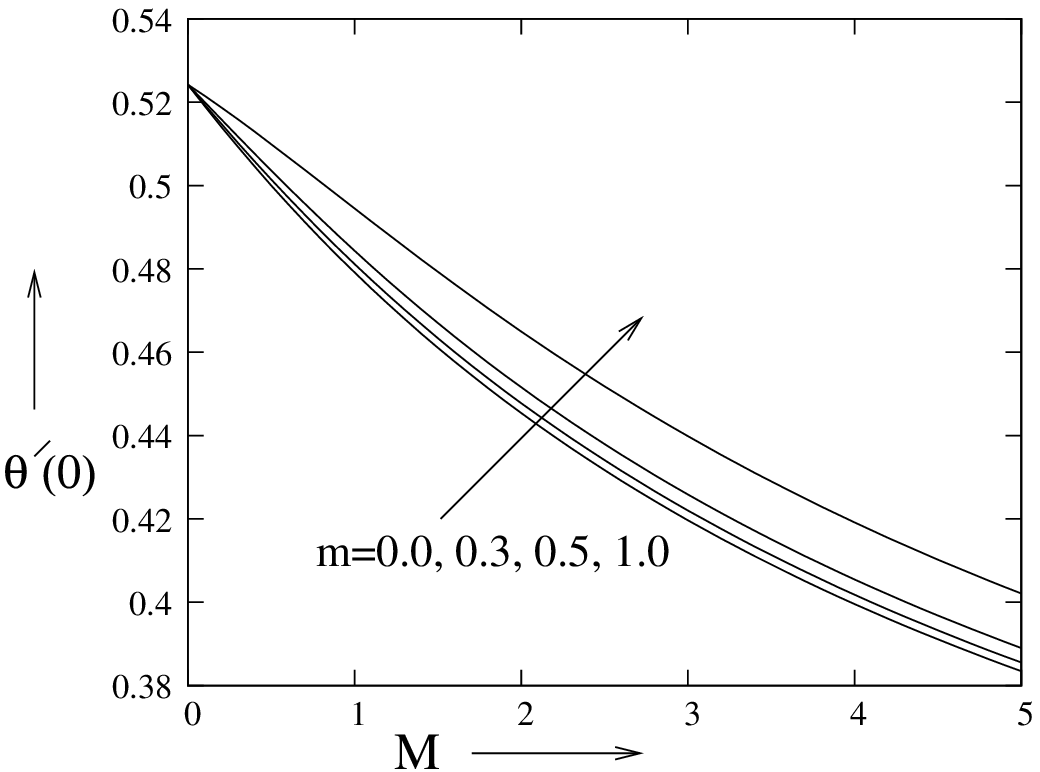}\\
      Fig. 17 Distribution of heat transfer rate at the channel wall for
different values of Hall parameter
     $m$ and $Pr=0.7,~ Gr=Gc=0.5, ~Sc=0.5, ~n=1, ~\gamma =0.1 $ \\

     \includegraphics[width=3.8in,height=3.0in ]{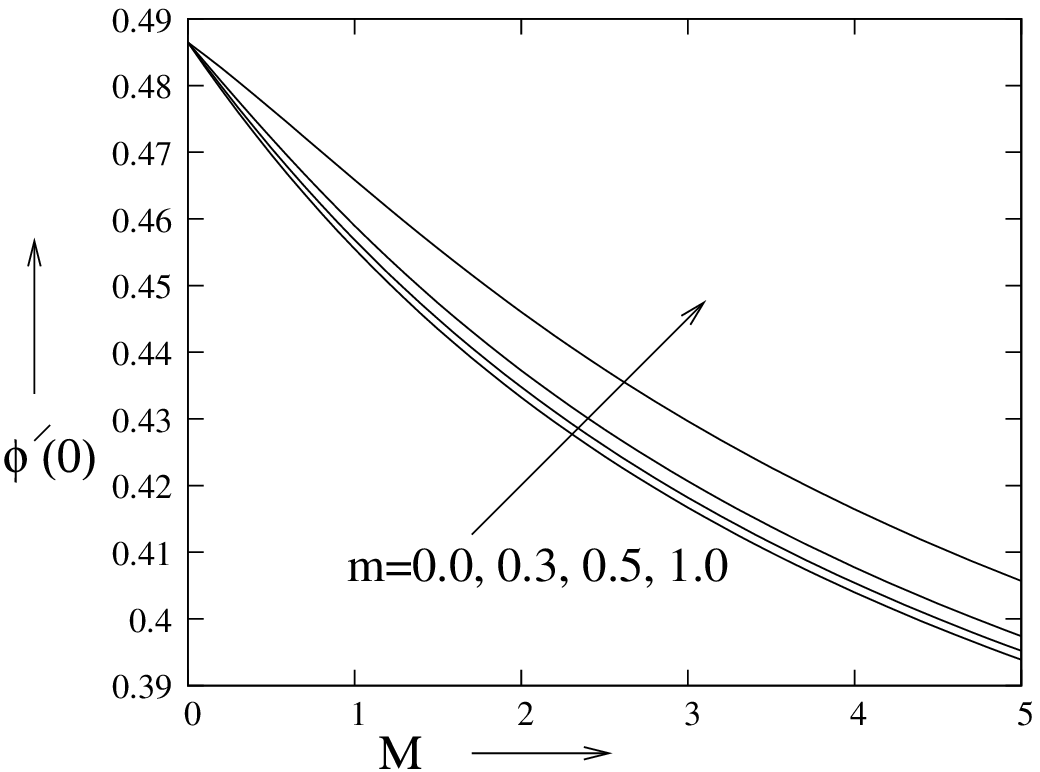}\\
     Fig. 18 Distribution of the rate of concentration at the channel wall for
different values of Hall parameter
     $m$ and $Pr=0.7,~ Gr=Gc=0.5, ~Sc=0.5, ~n=1, ~\gamma =0.1 $ \\
\end{center}
\end{minipage}\vspace*{.5cm}\\
\end{document}